\documentclass[conference]{main}
\IEEEoverridecommandlockouts
\usepackage{algorithm}
\usepackage{algpseudocode}
\usepackage{caption}
\usepackage[bottom]{footmisc}
\usepackage{multicol}
\usepackage{graphicx}
\usepackage{url}
\usepackage{xcolor}
\usepackage{float}
\usepackage{amsmath}
\usepackage{amsfonts}
\usepackage{amssymb}
\usepackage{amsthm}
\usepackage{mathrsfs}
\usepackage{amstext}
\usepackage{commath}
\usepackage[noadjust]{cite}
\usepackage{url}
\usepackage[all]{xy}
\usepackage{epsfig}
\usepackage{epstopdf}
\usepackage{tabularx}
\usepackage{graphicx}
\usepackage{graphics}
\usepackage{float}
\usepackage{multicol}
\usepackage{makeidx}
\usepackage{accents}
\usepackage[mathscr]{euscript}
\usepackage[utf8x]{inputenc}
\usepackage{algorithm}
\usepackage{algpseudocode}
\usepackage{comment}
\usepackage{makecell}
\usepackage{adjustbox}
\usepackage{multirow}
\usepackage{steinmetz}
\usepackage{rotating}
\usepackage{soul}
\usepackage{balance}
\usepackage{enumitem}
\usepackage{pifont}
\usepackage{amssymb}
\usepackage[acronym]{glossaries}
\usepackage{multirow}
\IEEEoverridecommandlockouts
\newacronym{ami}{AMI}{Advanced Metering Infrastructure}
\newacronym{bb}{BB}{Broadband}
\newacronym{bis}{BIS}{Broadband Impedance Spectroscopy}
\newacronym{cenelec}{CENELEC}{European Committee for Electrotechnical Standardization}
\newacronym{cfr}{CFR}{Channel Frequency Response}
\newacronym{cir}{CIR}{Channel Impulse Response}
\newacronym{ctdr}{CTDR}{Correlation Time Domain Reflectometry}
\newacronym{cs}{CS}{Compressed Sensing}
\newacronym{ctf}{CTF}{Channel Transfer Function}
\newacronym{csi}{CSI}{Channel State Information}
\newacronym{der}{DER}{Distributed Energy Resource}
\newacronym{dnra}{DNRA}{Dynamic Network Reconstruction Algorithm}
\newacronym{eis}{EIS}{Electrical Impedance Spectroscopy}
\newacronym{fdr}{FDR}{Frequency Domain Reflectometry}
\newacronym{fft}{FFT}{Fast Fourier Transform}
\newacronym{fmcw}{FMCW}{Frequency Modulated Carrier Wave}
\newacronym{glrt}{GLRT}{Generalized Likelihood Ratio Test}
\newacronym{hif}{HIF}{High Impedance Fault}
\newacronym{icmp}{ICMP}{Internet Control Message Protocol}
\newacronym{ift}{IFT}{Inverse Fourier Transform}
\newacronym{iot}{IoT}{Internet of Things}
\newacronym{it}{IT}{Information Technology}
\newacronym{jtfdr}{JTFDR}{Joint Time Frequency Domain Reflectometry }
\newacronym{lv}{LV}{Low Voltage}
\newacronym{mg}{MG}{Microgrid}
\newacronym{ml}{ML}{Machine Learning}
\newacronym{mse}{MSE}{Mean Square Error}
\newacronym{mtl}{MTL}{Multiconductor Transmission Line}
\newacronym{nn}{NN}{Neural Networks}
\newacronym{nns}{NNs}{Neural Networks}
\newacronym{ntp}{NTP}{Network Time Protocol}
\newacronym{ofdm}{OFDM}{Orthogonal Frequency Division Multiplexing}
\newacronym{ot}{OT}{Operation Technology}
\newacronym{pd-fdr}{PD-FDR}{Phase Detection-FDR}
\newacronym{phev}{PHEV}{ Plug-in Hybrid Electric Vehicle}
\newacronym{phy}{PHY}{Physical Layer}
\newacronym{pilc}{PILC}{Paper-Insulated Lead-Covered}
\newacronym{plc}{PLC}{Power Line Communication}
\newacronym{plm}{PLM}{Power Line Modem}
\newacronym{plms}{PLMs}{Power Line Modems}
\newacronym{pls}{PLS}{Physical Layer Security}
\newacronym{pul}{PUL}{Per Unit Length}
\newacronym{p2p}{P2P}{Point-to-Point}
\newacronym{rnja}{RNJA}{Rooted Neighbor-Joining Algorithm}
\newacronym{sgs}{SGs}{Smart Grids}
\newacronym{snr}{SNR}{Signal to Noise Ratio}
\newacronym{sp}{SP}{Single-Point}
\newacronym{stp}{STP}{Shortest Time Path}
\newacronym{svm}{SVM}{Support Vector Machine}
\newacronym{svms}{SVMs}{Support Vector Machines}
\newacronym{swr}{SWR-FDR}{Standing Wave Ratio-FDR}
\newacronym{tdr}{TDR}{Time Domain Reflectometry}
\newacronym{tem}{TEM}{Transverse Electromagnetic}
\newacronym{tf}{TF}{Transfer Function}
\newacronym{tfdr}{TFDR}{Time Frequency Domain Reflectometry}
\newacronym{toa}{ToA}{Time of Arrival}
\newacronym{tof}{ToF}{Time of Flight}
\newacronym{ttl}{TTL}{Time to Live}
\newacronym{unb}{UNB}{Ultra Narrow Band}
\newacronym{wss}{WSS}{Wide Sense Symmetry}
\newacronym{w-fdr}{W-FDR}{Wideband-FDR}
\newacronym{wt}{WT}{Water Treeing}
\newacronym{xlpe}{XLPE}{Cross-Linked Polyethylene}
\newacronym{}{}{}
\newacronym{hs-ofdm}{HS-OFDM}{Hermi-tian Symmetric OFDM}
\newacronym{uwb}{UWB}{Ultra-Wideband}
\newacronym{css}{CSS}{Chirp Spread Spectrum}
\newacronym{emi}{EMI}{Electromagnetic Interferences}
\newacronym{tdoa}{TDoA}{Time Difference of Arrival}
\newacronym{plid}{PL-ID}{Power Line Identification}
\makeglossaries
\begin{document}
\raggedbottom

\title{Inferring Power Grid Information with Power Line Communications: Review and Insights}
%
\author{
Javier Hernandez Fernandez$^{1}$, Abdulah Jarouf$^{2}$, Aymen Omri$^1$, and Roberto Di Pietro$^3$\\
$^1$Iberdrola Innovation Middle East, Doha, Qatar.\\
$^2$Hamad Bin Khalifa University, College of Science and Engineering, \\ Division of Information and Computing Technology, Doha, Qatar.\\
$^3$King Abdullah Univrsity of Science and Technology, CEMSE, RC3, Thuwal, Saudi Arabia.
}

\maketitle
\begin{abstract}
High-frequency signals were widely studied in the last decade to identify grid and channel conditions in power lines. \acrfull{plms} operating on the grid's physical layer are capable of transmitting such signals to infer information about the medium. When applied to the electrical grid, one of the key advantages of \acrfull{plc} is its capacity to use signals to provide information about the grid itself. This makes PLC an ideal communication technology for smart grid applications, particularly in the realms of grid monitoring and surveillance.
In this paper, we focus on PLC grid information inference and provide several contributions: a classification of PLC-based applications, a review of the relevant literature, and insights to further advance the field. Our research identified contributions addressing PLMs for three main grid information inference applications: topology inference, anomaly detection, and grid cybersecurity. We utilize the outcome of our review to shed light on the current limitations of the research contributions and suggest future research directions in this field.
\end{abstract}


\section{Introduction} 
\IEEEPARstart{P}{ower} Line Communications (PLC) is a technology that transmits data over power lines and serves the grid for a wide range of applications \cite{7467440}. Some of the key uses of PLC include remote meter data collection, grid control, fault detection, and data transmissions \cite{YIGIT2014}. Recent advances in grid technologies, such as Microgrids or \cite{Yoldas2017EnhancingSGwithMIC} \acrfull{der}s \cite{AGUERO2018}, have highlighted the need for enhanced methods for grid protection, communication, and control \cite{machowski2020power,1519722,5503917}. These methods primarily rely on automating grid operations to maintain generation-demand balance and stability \cite{SAIDU20223192}. Additionally, they enable continuous monitoring of grid assets, facilitate two-way communication, and improve demand-side management \cite{Guelpa2021DRSurvey}. Monitoring grid assets, such as lines and nodes, typically involves deploying measurement equipment like sensors around the grid to gather diagnostic data \cite{RIVAS2020,Kababji2024DesignPLC}.
A significant advantage of PLC signals is their ability to simultaneously provide information about the grid and address communication needs, with some areas of research focusing on enhancing grid monitoring solutions using this technology \cite{passerini2017NetworkSensing}. 
Furthermore, security analyses of grid communications have demonstrated potential enhancements through the use of \acrfull{pls}-based PLC techniques \cite{HernandezFernandez2023performancePLS}. For instance, channel characteristics can be used to extract common randomness between grid nodes to derive shared security keys \cite{Henkel2020}. Additionally, variations in Channel State Information (CSI) and noise can provide insights into possible network intrusions and help identify their locations \cite{HernandezFernandez2023}.

Existing reviews of PLC technology have predominantly concentrated on the communications aspect or specific applications. To the best of our knowledge, there are no comprehensive reviews that focus on the grid information inference component. Two related surveys were found to cover specific topics such as the classification of cable fault detection methods in power line networks \cite{Furse2006, Shi2010}. Additionally, one review provided a general overview of PLC grid monitoring applications and methods \cite{Passerini2018}, and another focused on security \cite{PLCSurvey}.

\subsection{Contribution}
In light of the aforementioned limitations, this paper examines contributions related to PLC grid information inference and offers a classification of applications and methodologies. Specifically, we provide several contributions: 

\begin{itemize}
    \item \textcolor{black}{A categorization of PLC applications utilizing grid inferred information  organized into two primary domains: grid monitoring and cybersecurity applications.} 
    \item \textcolor{black}{An evaluation of the methodologies and techniques employed for each application, including their inputs, evaluation metrics, setup, advantages, and disadvantages.}
    \item We finally provide a general evaluation of the contributions in this field and present potential future research directions.
\end{itemize}
This study's results are expected to potentially influence this research community by organizing the current contributions, identifying the limitations, and providing ideas for future research directions. 

\begin{figure*}[!h]
    \centering
    \includegraphics[trim={0 1cm 0 1cm},clip]{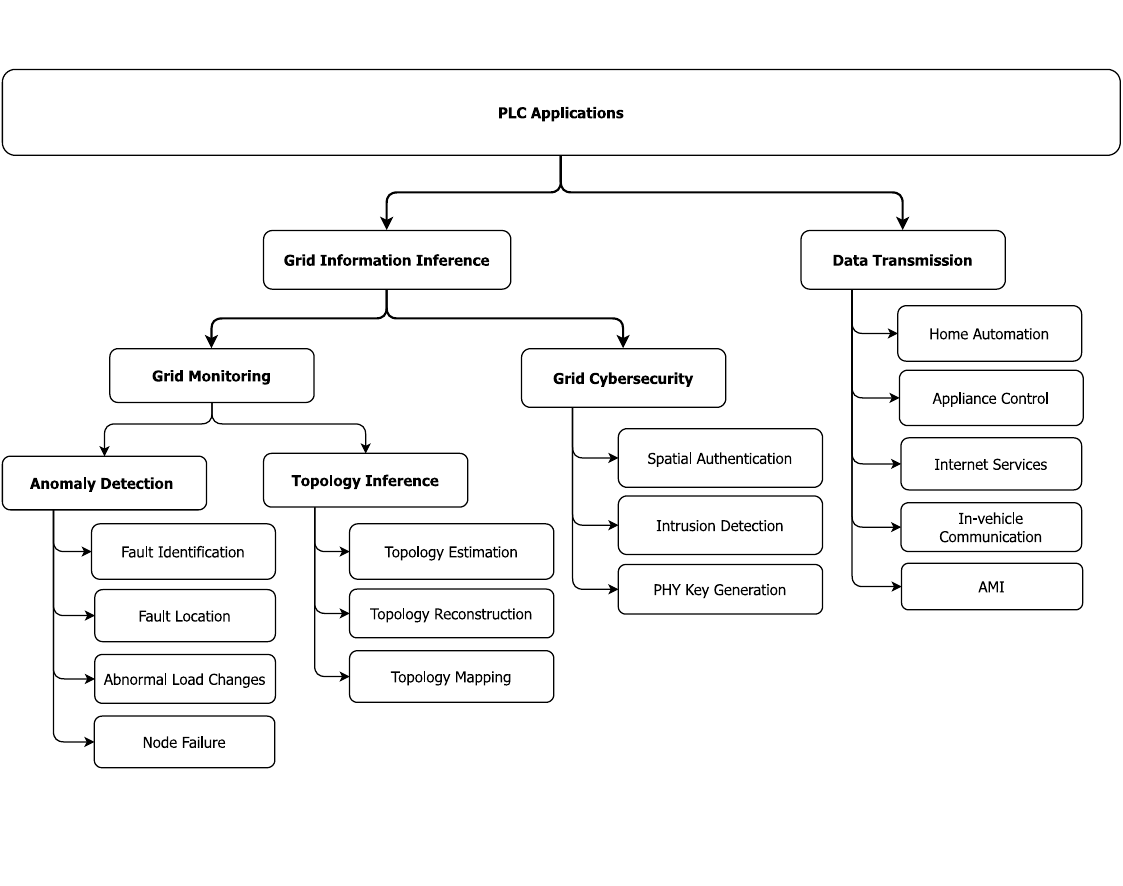}
    \caption{Classification of PLC applications.}
    \label{fig:applications}
\end{figure*}

\subsection{Paper Organization}

\textcolor{black}{The remainder of the paper begins with the proposed structure of PLC applications and provides some background information in Section \ref{background}. A review of the related contributions is presented in Section \ref{review1}. This is followed by the insights derived from the study and future research directions in Section \ref{discussion}, concluding with Section \ref{conclusion}.}

\section{Classification of PLC Applications}\label{background}
\textcolor{black}{Beyond serving as a communication backbone, PLC is also able to sense and infer latent grid information \cite{Galli2011}, therefore distinctly categorizing its applications into data transmission and grid information inference.} 

\textcolor{black}{PLC integrates measurement and communication functions through Power Line Modems (PLMs), offering an efficient solution for various applications. It enables real-time data transmission using existing electrical wiring to enhance efficiency and functionality. Over time, the development of PLMs has significantly improved, and nowadays, a variety of customizable modems exist that adapt to a wide range of applications and scenarios. They achieve this by combining different bandwidths, transmission schemes, and carrier frequencies depending on their intended use. Typically, PLMs are classified based on the frequency range in which they operate into three categories \cite{PLSinPLC_Book_2024}:}

\begin{itemize}
\item \textcolor{black}{Ultra Narrowband PLC (UNB-PLC) operates in low-frequency range between 0.3 kHz and 3 kHz and  supports very low data rates of around 100 bits per second (bps), but can achieve extremely long transmission distances. Therefore UNB-PLC is limited to  control applications requiring long-range communication, such as distribution automation and certain IoT applications.} 
\item \textcolor{black}{Narrowband PLC (NB-PLC) uses the frequency range below 500 kHz and supports data rates of kilobits per second (Kbps). It is normally used for applications such as smart metering or grid monitoring. }
\item \textcolor{black}{Broadband PLC (BB-PLC) works in the frequency range above 1.8 MHz, typically between 2 MHz and 30 MHz, and can achieve data rates of megabits per second (Mbps). BB-PLC is suitable for applications requiring high-bandwidth but suffers from a shorter transmission range and is more susceptible noise and attenuation.}
\end{itemize}



\subsection{PLC for Data Transmission}
PLC for data transmission provides reliable and efficient communication services across all power grid levels, supporting various applications, including the implementation of \acrfull{ami} and home automation \cite{hallack2016PLChomeandIndch7}. Additionally, BB-PLC networks enable utilities to offer internet services and integrate with other communication networks, while also being considered for in-vehicle telecommunication and control applications \cite{pinto2015,antoniali2011,degardin2013}.

\subsection{PLC for Grid Information Inference}
Since the transmission medium spans the entire grid, PLMs can act as sensors by generating signals of various frequency ranges and types \cite{SHARMA2017,ergodic2022}. These measurements can be analyzed to extract information about the grid's physical layer, with applications broadly classified into grid monitoring and security. A detailed discussion of these applications is provided in the following sections and summarized in Fig. \ref{fig:applications}. 

\subsubsection{Grid Monitoring} 
The PLC channel is time-variant due to changes in grid physical parameters like line impedance, loads, and topology \cite{en13123098}, which impact grid operations such as communication, control, and asset management \cite{Liang2019, Lehmann2016}. We present topology inference and anomaly detection as the primary applications of grid monitoring. Topology inference involves deducing grid parameters from state measurements and includes topology estimation, reconstruction, and mapping. Topology estimation predicts the grid’s arrangement, topology reconstruction recalculates parameters with prior knowledge, and topology mapping assigns assets to specific grid sectors \cite{Pappu2018, Diwold2015}. Anomaly detection identifies, classifies, and localizes abnormal grid conditions, addressing issues like line faults, broken conductors, and node failures.  \textcolor{black}{Most existing solutions require dedicated equipment or interfere with standard operations \cite{Huo2019}, but PLC can mitigate these drawbacks through continuous evaluation using PLMs. Current anomaly detection in Power Line Communication (PLC) primarily targets line faults.}

As depicted in Fig. \ref{fig:applications} applications of anomaly detection include fault identification, fault location, abnormal load changes, and node failure. 
Fault identification involves detecting and classifying the nature of faults \cite{Milioudis2012a, Milioudis2015}, as well as issues related to degradation and aging \cite{Huo2018a, Huo2018}, or more severe faults \cite{Passerini2019a}, that could impact the power transmissions. Fault location determines the fault position in
the network \cite{Huo2019, Passerini2019}. The location could be given as a distance from PLMs or as a particular line or branch in the network topology, depending on the equipment and algorithm used. \textcolor{black}{Abnormal load changes refer to deviations from typical load patterns \cite{LABRADORRIVAS2020106602}, that could indicate potential issues within grid. Node failure, on the other hand, focuses on the operational status of the network’s devices \cite{Rao2011}}.

\subsubsection{Grid Cybersecurity} 
Security in power systems has gained significant attention recently as the grid digitizes and incorporates more advanced technologies. PLC systems face physical and cyber threats, including unauthorized access, data breaches, and electromagnetic interference. \textcolor{black}{Traditional cryptography methods are used but may be ineffective in computationally constrained devices \cite{Huo2018a}.}   
Therefore, adopting information-theoretic security algorithms could complement existing ones to provide a more strict notion of security \cite{10.1007/978-3-319-23609-4_8}. The field of \acrshort{pls} in PLC systems is addressing this gap by investigating techniques such as intrusion detection, spatial authentication or key generation \cite{PLSinPLC_Book_2024}. 

\section{Review of PLC grid information inference} \label{review1}
This section presents the state-of-art of grid monitoring and cybersecurity fields of PLC grid information inference. The contributions are presented in three subsections: topology inference, anomaly detection, and grid cybersecurity. 

\subsection{Topology Inference}
Topology inference involves deducing topology parameters from the grid state and can be generally categorized into physical, network, and application layer-based estimation approaches. Table \ref{topology_sum} summarizes the contributions and their features, while Table \ref{topology_algs} compares the advantages and limitations of the inferring algorithms."

\begin{table*}[]
\centering
\caption{Summary of Research Contributions for Topology Inference }
\label{topology_sum}
\setlength\extrarowheight{4pt}
\begin{tabular}{|c| m{0.09\linewidth}|>\centering m{0.07\linewidth}|>\centering m{0.1\linewidth}|>\centering m{0.03\linewidth}|>\centering m{0.11\linewidth}|m{0.3\linewidth}|}
\hline
\textbf{References} & \textbf{Application}                              & \textbf{Layer} &\textbf{Measurement Technique} & \textbf{CSI} & \textbf{Setup} & \textbf{\* \* \* \* \* \* \* \* \* \* \* \* \* \* \* \* \* \* \* \* Summary}                                                                                                                                                                                                                                                            \\ \hline
 \cite{Ahmed2012}            & \multirow[c]{6}{0.1\linewidth}[-1 in]{Topology Estimation} &  PHY                       & FDR                            & \ding{51}          & Single-Point     & CFR was obtained via signal reflections correlation, CIR was then used to calculate the estimated discontinuity distances based on peak occurrence time. The graph is estimated via Peak-by-peak processing.                                      \vspace{0.04in}                 \\ \cline{1-1} \cline{3-7} 
\cite{Zhang2016}           &                                          &  PHY                       & TFDR                     & \ding{51}          & Single-Point      & TFDR is used to measure discontinuity distances with high resolution ($1m$) and less complexity. The distances are fed to a node-by-node algorithm to predict the graph.           \vspace{0.03in}                                                                                \\ \cline{1-1} \cline{3-7} 
\cite{Ulrich2015}           &                                          &    PHY                       & FDR                            & \ding{55}            & Extended-Single-Point      & CoMaTeCh algorithm runs on a central node to predict topology parameters and graphs based on the FDR distance estimation of multiple PLC nodes.                                                                                                                    \vspace{0.04in}   \\ \cline{1-1} \cline{3-7} 
\cite{Ahmed2013, Ma2015}           &                                          &  PHY                       & OFDM Pilots                    & \ding{51}          & Extended-P2P       & OFDM pilot transmission of all node pairs are used to estimate the signal time of arrival on a parametric estimation basis. RNJA and DNRA predict the topologies in \cite{Ahmed2013} and \cite{Ma2015}, respectively. \vspace{0.04in}   \\ \cline{1-1} \cline{3-7} 
\cite{Lampe2013, Erseghe2013}            &                                          &     Network                   & NTP Handshake                  & \ding{55}            & Extended-P2P      & NTP handshake protocol is used as the distance measurement technique based on ToA. While \cite{Lampe2013} uses RNJA, \cite{Erseghe2013} uses a hypothesis testing approach based on GLRT for graph prediction.
\vspace{-0.1in}
\vspace{0.00001in} \\

\cline{1-1} \cline{3-7} 
\cite{Passerini2017}           &                                          &     PHY                   & Impedance measurements                  & \ding{55}            & Extended-Single-Point      & Based on the prior knowledge of load and cable parameters, the admittance is measured at all node ends. The topology graph and line length are then derived correctly in robust networks.                    \vspace{0.04in}   \\ \hline

 \cite{Aouichak2017}           & \multirow[c]{3}{0.1\linewidth}[-0.5 in]{Topology Reconstruction} &  PHY                      & ToF            & \ding{55}              & Extended-P2P       &     Propagation delays between indoor network outlets are collected using a pulse emitter and receiver. The outlet pair distances are then calculated based on matrix equations and prior grid information.                                                                                                                                                                                                                                                               \vspace{0.04in}   \\ \cline{1-1} \cline{3-7} 
\cite{Costabeber2011}            &                                          & PHY                      & ToF            & \ding{55}              & Extended-P2P    &     Distance ranging was approached based on the two-way handshake protocol to reconstruct the address-distance pairs of all nodes. The node neighbors are calculated based on the shortest direct paths.                                                                                                                                                                                                                                                               \vspace{0.04in}   \\ \cline{1-1} \cline{3-7} 
\cite{Wang2019}           &                                          & Network                & Traceroute                     & \ding{55}            & Extended-P2P &     Traceroute protocol is used to discover the message hubs between two nodes. The collected nodes' responses helped reconstruct the topology and predict the node connections.                                                                                                                                                                                                                                                                 \vspace{0.04in}   \\ \hline
\cite{Marron2013}           & \multirow[c]{2}{0.1\linewidth}[-0.2 in]{Topology Mapping}        &Network                & Keep-Alive / KCL                     & \ding{55}    &      Extended-P2P  & PLC Keep-Alive Packets and current RMS are used to map smart meters to their feeders. The PLC topology is used as a lower-priority source of information.                                                                                                                                                                                                                                                            \vspace{0.04in}   \\ \cline{1-1} \cline{3-7} 
\cite{Pappu2018, Diwold2015}           &                                          &    Application    &  Energy measurements. & \ding{55}             & Multi-Point & Voltage measurements collected from smart meters were used for feeder identification and grid mapping. Meter grouping was approached by signal correlation of the meter measurements.   \vspace{0.04in}   \\ \hline
\end{tabular}%
\end{table*}

\begin{table*}[!]
\centering
\caption{Summary of Topology Estimation Algorithms}
\label{topology_algs}
\setlength\extrarowheight{4pt}
\begin{tabular}{|>\centering m{0.1\linewidth}|>\centering m{0.09\linewidth}|>\centering m{0.08\linewidth}|>\centering m{0.1\linewidth}|>\centering m{0.1\linewidth}|>\centering m{0.15\linewidth}|m{0.2\linewidth}|}
\hline
\textbf{Algorithm}         & \textbf{Input} & \textbf{Output} & \textbf{Convergence Ratio} & \textbf{Evaluation Metric} & \textbf{Advantages} & \textbf{\* \* \* Disadvantages/Limitations}\\ \hline
Peak-by-peak \cite{Ahmed2012}       & 
\begin{itemize}[leftmargin=*] 
    \item CIR peaks 
\end{itemize}
& A set of possible topologies & Higher for longer measurement ranges &
FDR consistency&
\begin{itemize}[leftmargin=*]
    \item No prior grid information is required.
\end{itemize}
&
\begin{itemize}[leftmargin=*]
        \item Complexity grows exponentially with the number of nodes.
    \item Includes redundant search of branches. \vspace{-0.12in}
\end{itemize} 
\\ \hline
Node-by-node \cite{Zhang2016}       & 
\begin{itemize}[leftmargin=*]
    \item $N_{max}$
    \item CIR peaks 
\end{itemize} & A set of possible topologies & Higher for smaller $N_{max}$ and distances & Path lengths consistency with peak locations & 
\begin{itemize}[leftmargin=*]
    
    \item Ambiguity rate is $5.6\%$ in the worst case.
\end{itemize}
&
\begin{itemize}[leftmargin=*]
    \item Requires prior grid information: $N_{max}$.
    \item Works only for indoor and similar PLC networks. \vspace{-0.12in}
\end{itemize}
\\ \hline
RNJA \cite{Ahmed2013} \cite{Lampe2013}       & 
\begin{itemize}[leftmargin=*]
    \item Distances 
    \item $N$ 
\end{itemize}
& A single binary tree topology & High for robust measurement methods& Distance from root node& 
\begin{itemize}[leftmargin=*]
    \item A single solution is provided.
\end{itemize}
&
\begin{itemize}[leftmargin=*]
    \item Requires prior grid information: topology type and number of nodes.
    \item Works only for tree topology (binary trees). \vspace{-0.12in}
\end{itemize}
\\ \hline
\textcolor{black}{Peak-by-peak \cite{Mlynek2016}}           & 
\begin{itemize}[leftmargin=*]
    \item \textcolor{black}{Multi-node
CIR}
\end{itemize}
& \textcolor{black}{A single tree topology} & \textcolor{black}{High for robust
measurement methods} & \textcolor{black}{CIR consistency} &
\begin{itemize}[leftmargin=*]

    \item \textcolor{black}{The method does not require a PLC device at each point of the tested topology.}
\end{itemize}
&
\begin{itemize}[leftmargin=*]
        \item \textcolor{black}{Sensitive to the PLC CIR estimation error.}
\end{itemize}
\\ \hline
CoMaTeCh \cite{Ulrich2015}           & 
\begin{itemize}[leftmargin=*]
    \item Multi-node CIR 
\end{itemize}
& A single tree topology & Higher for higher number of measurement points & CIR consistency &
\begin{itemize}[leftmargin=*]

    \item Few nodes are required to be PLC-equipped.
\end{itemize}
&
\begin{itemize}[leftmargin=*]
        \item Works only for tree topology.
\end{itemize}
\\ \hline
Hypothesis testing \cite{Erseghe2013} &
\begin{itemize}[leftmargin=*] 
    \item Shortest Time Paths (STPs) 
\end{itemize}
& A binary neighborhood matrix for all nodes &     High for robust measurement methods         & Likelihood probability     &
\begin{itemize}[leftmargin=*]
    \item A-GLRT provides simplifications of F-GLRT for online deployments.
\end{itemize}
&
\begin{itemize}[leftmargin=*]
                \item Does not provide graph or node connections. 
\end{itemize}
\\ \hline
DNRA \cite{Ma2015}            & 
\begin{itemize}[leftmargin=*]
\item Distances
\item Root node 
\item $N$
\item Electrical length resolution    \vspace{-0.11in}
\end{itemize}
& A single tree topology  & High for robust measurement methods & Distance from root node &
\begin{itemize}[leftmargin=*]
   \item Less complexity ($O(N^2)$) than RNJA ($O(N^3)$). 
   \item A single solution is provided. \vspace{-0.15in}
\end{itemize}
&
\begin{itemize}[leftmargin=*]
        \item Requires prior grid information: topology type and number of nodes.
        \item Works only for tree topology and targets outdoor SMGs. \vspace{-0.15in}
\end{itemize}
\\ \hline
\end{tabular}
\end{table*}
\subsubsection{Physical Layer Topology Inference}

Using a simple setup to blindly infer topology parameters and node connections, Ahmed \emph{et al.} proposed an algorithm to estimate the PLN network topology based on distance measurements collected via \acrfull{fdr} \cite{Ahmed2012}. The proposed technique correlated the reflections of multiple frequency-shifted signals over a specific measurement bandwidth to reconstruct the Channel Frequency Response (CFR). Consequently, distance measurements were obtained from the \acrfull{cir} using the \acrfull{ift} and the signal propagation speed. CIR produces a final graph consisting of peaks at discontinuity locations. As for the graph prediction algorithm, peaks of the CIR plot are processed individually regardless of the peak magnitude or phase to form a list of candidate topologies at each peak processing time (iteration). Based on the consistency of the measured CIR and every candidate topology's simulated CIR, topologies are either accepted for the next iteration or rejected. However, since the complexity of this peak-by-peak algorithm grows exponentially with every impulse detected, and the resolution measurement of the FDR method is kept relatively low ($20 m$), Zhang \emph{et al.} proposed a more efficient algorithm based on a higher resolution distance measurements \cite{Zhang2016}. Distances are estimated based on a joint \acrfull{tfdr} approach which provides a much higher resolution ($1 m$) with a reduced measurement bandwidth. Like the peak-by-peak algorithm, discontinuity locations are only considered for the inference algorithm operating node-by-node. For this algorithm to run successfully, the expected maximum number of nodes in the network must be provided in advance. However, the greedy node-by-node search algorithm follows a similar concept to the peak-by-peak approach since it compares the theoretical path lengths of the candidate set to the measured ones iteratively as an evaluation metric. Ravishankar \emph{et al.} combined \acrfull{ctdr} and FDR in a hybrid method to estimate the bus topology. Multiple measurement points are required to obtain the CTDR response, revealing the initial approximate topology. The genetic algorithm (GA) was then used to refine the estimation by the \acrfull{mse} comparison of computed and measured FDR responses of the estimated and actual topology, respectively \cite{Ravishankar2016, Ravishankar2012LineTE}.

\textcolor{black}{Mlynek \emph{et al.} \cite{Mlynek2016} introduced and investigated a CIR-based method for managing a line 
topology, by estimating the topological parameter, e.g., number of branches, branched line length, and direct line length of unknown topology.}
In contrast, the work in \cite{Ulrich2015} uses FDR given its higher signal-to-noise ratio and resolution. The proposed algorithm, (CoMaTeCh) is used to infer network topologies using reflection measurements from multiple cable ends and does not require previous knowledge of the network. A drawback of this method is that the performance is highly dependent on the number of measurement points. With too few measurements, there is a risk of accidental reflection matches, which can result in incorrect results. 

Passerini \emph{et al.} investigated the possibility of topology estimation based purely on admittance measurements  \cite{Passerini2017}. The model solution relies on transmission line theory to correlate the line length connecting two nodes with the network and load admittance measurements at the two ends. The algorithm targets the topology estimation in power line networks without noise, yet further analysis was done to understand the impact of noise on the algorithm. However, since this method does not guarantee an online measurement of the load's admittance, it is advised to use FDR techniques or other real-time impedance estimation methods such as the one in \cite{Liang2019}. 
Aouichal \emph{et al.} deployed multiple pulse emitters and receivers inside an indoor network to measure the distances between power outlets based on the propagation delay of impulses. Although the resolution obtained is high, this method requires a number of measurements proportional to the network’s size and requires some prior information on the topology \cite{Aouichak2017}. 
ToF was also approached by Costabeber \emph{et al.} to compute the distances between nodes to reconstruct each node's local topology based on shortest direct paths \cite{Costabeber2011}. This ranging approach was part of a token-based distributed generation control model to deliver power efficiently to loads. 

In \cite{Ahmed2012a}, the authors presented and discussed both parametric and non-parametric methods for distance estimation using PLC and reflectometry techniques. Subsequently, Ahmed \emph{et al.} used the frequency-shifted time-delayed \acrfull{ofdm} samples to estimate \acrfull{toa} based on CFR derivations \cite{Ahmed2013}. The same measurement method was approached in \cite{Ma2015}, with a different choice of the parameter estimation algorithm. Ahmed \emph{et al.} used the subspace method to obtain the missing ToA parameters. At the same time, \cite{Ma2015} used the Bayesian \acrfull{cs} method to improve the estimation under high \acrfull{snr} conditions and to increase the spectrum efficiency with reduced pilot overhead. In both cases, The estimated end-to-end distances are transmitted to a central node that runs the topology estimation algorithm. Considering a tree topology, the \acrfull{rnja} was used in \cite{Ahmed2013} to combine the leaf nodes based on the reported distances until one final leaf node remains unlinked. \acrfull{dnra} \cite{Ma2015} works similarly but without bulk-inputting all the node-pair measurements. However, both algorithms have a good convergence rate and tolerance to the error in path length estimation. 

\subsubsection{Network Layer Topology Inference}
Network protocols such as \acrfull{ntp} and traceroute can infer topology parameters based on signal propagation time estimation. Previous contributions utilized NTP to estimate the \acrfull{stp} between two nodes in moderate to high SNR environments to have a high detection accuracy of ToA \cite{Lampe2013, Erseghe2013}. 
Consequently, Erseghe \emph{ et al.} followed a hypothesis-based testing method to decide node connections to neighboring nodes relying on the Generalized Likelihood Ratio Test (GLRT) to strengthen the correct hypothesis selection rate \cite{Erseghe2013}. 
Likewise, Ahmed \emph{et al.} used the estimated \acrshort{stp} measurements to input the RNJA algorithm. Most importantly, with the end-to-end measurements, it can now detect and estimate the distance of a PLC-unequipped node in the middle of two PLC-equipped nodes. 
In a related approach, \cite{Wang2019} implemented the topology reconstruction algorithm based on traceroute data collected from distributed PLC nodes. The traceroute data include the \acrfull{icmp} command and \acrfull{ttl} values to identify routers on the path. The topology reconstruction algorithm operates in three phases: classifying nodes in a virtual topology, generating merging options, and iteratively connecting links until the set of options is null, resulting in the final topology.

The PLC Keep-alive packets are used in \cite{Marron2013} to map feeders to smart meters. The Root Mean Square (RMS) current of each signal was extracted using an inductive coupler at each feeder. Statistical analysis is then applied to map the PLC devices to their respective feeders. The underlying communication topology followed by the PLC is used as an alternative source of information. The primary benefit of this solution is that it reuses the available information provided by the PLC smart meters. 

\subsubsection{Application Layer Topology Inference}
Application layer data processing based on meter statistical measurements, such as energy and voltage measurements, showed potential success in topology inference. Meter energy measurements were used in \cite{Pappu2018} for phase nodes and topology identification based on the energy conservation concepts in a distributed network. 
Voltage measurements of many meters were correlated and analyzed to provide node mapping to topology sections and discover feeder assignment \cite{Diwold2015}.

\begin{table*}[]
\centering
\caption{Summary of Research Contributions for Anomaly Detection}
\label{anomaly_sum}
\setlength\extrarowheight{4pt}
\begin{tabular}{|c|>\centering m{0.07\linewidth}|>\centering m{0.1\linewidth}|>\centering m{0.05\linewidth}|>\centering m{0.06\linewidth}|>\centering m{0.1\linewidth}|m{0.32\linewidth}|} \hline
\textbf{Reference}                                                 & \textbf{Fault Type}  & \textbf{Classification}                                    & \textbf{Severity} & \textbf{Location} & \textbf{Measurements}     & \textbf{\* \* \* \* \*  \* \* \* \* \* \* \* \* \* \* \* \* \* \* \* Summary}                                                                                                                                                                                                            \\ \hline
\cite{Milioudis2010} \cite{Milioudis2011} \cite{Milioudis2012} & \multirow[c]{3}{0.45\linewidth}[-0.5 in]{HIF} & \ding{55}                                                  & \ding{55}         & \ding{55}         & Input impedance      & Detection of HIF in CENELEC A and BB-PLC bands based on the evaluation and analysis of the measured input impedance $Z_{in}$. A baseline model approach was followed for varying topology types and parameters.            \vspace{0.04in}   \\ \cline{1-1} \cline{3-7} 
\cite{Milioudis2012a}                                          &                      & HIF or broken conductor                                    & \ding{55}         & \ding{51}         & CIR                  & Reflectometric and E2E measurements were used to construct the CIR, which was compared to a reference CIR in a healthy state to determine the fault location.                                                               \vspace{0.04in}   \\ \cline{1-1} \cline{3-7} 
\cite{Milioudis2015}                                           &                      & \ding{55}                                                  & \ding{55}         & \ding{55}         & Input impedance      & Generalization of the work proposed in \cite{Milioudis2012} to cover the multi-conductor transmission lines configurations. \\ \cline{1-1} \cline{3-7} 
\textcolor{black}{\cite{DEOLIVEIRA2022107168}}                                          &                      & \textcolor{black}{\ding{55}}                                                  & \textcolor{black}{\ding{55}}         & \textcolor{black}{\ding{55}}         & \textcolor{black}{Time domain pulses, i.e., HS-OFDM, impulsive UWB, and CSS}      & \textcolor{black}{Typical PLC
Pulses were investigated to detect high-impedance faults, using time-domain reflectometry}                                                                                                   \vspace{0.04in}   \\ \hline
\cite{Forstel2017}\cite{Huo2018a}                              & WT                   & \multirow[c]{4}{\linewidth}[-0.8 in]{Localized, distributed, or load variation} & \ding{51}         & \ding{55}         & CFR                  & Water Treeing (WT) detection in XLPE cables based on a WT cable aging model and PLC CFR measurements. ML models enabled the classification and evaluation of fault severity and load variations.                           \vspace{0.04in}   \\ \cline{1-2} \cline{4-7} 
\cite{Huo2018}                                                 & Thermal              &                                                            & \ding{55}         & \ding{55}         & CFR                  & Cable degradation due to thermal conditions were modeled for PILC cables, and PLC CFR measurements were used to detect and classify faults in a similar manner to \cite{Huo2018a}.                                           \vspace{0.04in}   \\ \cline{1-2} \cline{4-7} 
\cite{Huo2019}                                                 & WT                   &                                                            & \ding{51}         & \ding{51}         & CIR \&         JTFDR & A complete ML framework for WT fault detection, classification, and localization was proposed based on the findings of \cite{Huo2018a}. The fault localization was done based on the CIR estimation of JTFDR.          \\ \cline{1-2} \cline{4-7} 
\textcolor{black}{\cite{Yang2013}}                                                & \textcolor{black}{WT}                  &                                                            & \textcolor{black}{\ding{51} }        & \textcolor{black}{\ding{51} }        & \textcolor{black}{CFR} & \textcolor{black}{Identification of different line faults, by investigating variations of CFR due to changes in power line parameters such as conductivity constant, relative permeability constant, and relative magnetic permeability} \\ \cline{1-2} \cline{4-7} 
\textcolor{black}{\cite{Huo2022}}                                                & \textcolor{black}{WT}                  &                                                            & \textcolor{black}{\ding{51} }        & \textcolor{black}{\ding{51} }        & \textcolor{black}{CSI} & \textcolor{black}{A power line sensing technique that can detect various types of cable anomalies without any prior domain knowledge is presented. The technique uses ML-based Time-series forecasting techniques to predict the PLC channel state information at any given point in time based on its historical data.} \vspace{0.04in}   \\ \cline{1-2} \cline{4-7}
\cite{Passerini2019a}\cite{Passerini2019}\cite{Lehmann2016}    & \multirow[c]{2}{0.45\linewidth}[-0.2 in]{N/A} &                          & \ding{55}         & \ding{51}         & CFR                  & The proposal of fault detection \cite{Passerini2019}\cite{Lehmann2016} and localization algorithm \cite{Lehmann2016} based on the chain representation model as given by the theoretical analysis in \cite{Passerini2019a}. \vspace{0.04in}   \\ \cline{1-1} \cline{3-7} 
\cite{Pinomaa2015}                                             &                      & \ding{55}                                                  & \ding{55}         & \ding{51}         & Input impedance      & Input impedance measurements are used to match the different fault properties obtained by the extensive search of parameters given in the line fault model.                                                                 \vspace{0.04in}   \\ \hline
\cite{10.3389/fenrg.2023.1103298}                                             &     Ground                 & \ding{55}                                                  & \ding{55}         & \ding{51}         & Impedance estimation via CFR     & CFR measurements were used to estimate the real-time cable impedance, followed by ML-based analysis to predict the cable condition.                                                                  \vspace{0.04in}   \\ \hline
\cite{https://doi.org/10.1049/smt2.12125}                                             &   Aging                   & \ding{55}                                                  & \ding{55}         & \ding{51}         & CTF      & Detection of aged cable segments based on the analysis of CTF measurements via advanced ML models.                        \vspace{0.04in}   \\ \hline

\textcolor{black}{\cite{Rao2011} }                                            &   \textcolor{black}{Connection problems}   & \textcolor{black}{\ding{55}}                                                  & \textcolor{black}{\ding{55}}         & \textcolor{black}{\ding{51}}         & \textcolor{black}{Smart Meter Signal Strength}       &    \textcolor{black}{Smart meter signal strength analysis were used to discover connection problems, e,g,, wrong node assignments, rapid signal level drop, and frequent signal level fluctuations.}                     \vspace{0.04in}   \\ \hline

\cite{ABDELKARIM2023}                                             &   Soft                   & \ding{55}                                                  & \ding{55}         & \ding{51}         & $T(f)$ estimation       & Online Transmission coefficient estimation was used to build a soft fault detection algorithm considering complex topology configurations and real-world data validation.                         \vspace{0.04in}   \\ \hline

\end{tabular}%
\end{table*}

\subsection{Anomaly Detection}
PLC-based anomaly detection contributions primarily focus on fault type and location identification and normally utilize physical layer techniques. Table \ref{anomaly_sum} summarizes the PLC anomaly detection contributions detailed below.

Milioudis \emph{et al.} studied the possibility of \acrfull{hif} detection using PLC signals in the CENELEC A, and BB-PLC bands \cite{Milioudis2010}. A detailed study was conducted to understand fault locations' effect and severity on the impedance frequency response. \textcolor{black}{In this context}, impedance frequency spectroscopy has been proposed a method to estimate the line impedance for frequency sweep signals. Estimating the impedance response is helpful for fault detection and possibly line discontinuity identification. \textcolor{black}{However, traditional Electrical Impedance Spectroscopy (EIS) is a time-consuming process, requiring multiple long period measurements to accurately estimate the impedance response of a system across a specific range of frequencies. This extended measurement
period may limit the efficiency of testing and delay data acquisition, particularly when rapid analysis is crucial.}\\
The HIF model was proposed as a parallel impedance element that affects the overall input impedance $Z_{in}$ measured at the end of the lines. Consequently, with $Z_{in}$ frequency response obtained in healthy conditions, HIF can be detected by the baseline model comparison method since HIF produces detectable changes to the frequency response. 
The work was further extended in \cite{Milioudis2011} to include complex topologies and account for the effects of ground resistivity and relative permittivity. 
Altogether, the efforts were combined in \cite{Milioudis2012} including practical considerations for implementation. Subsequently, a method for HIF location estimation was proposed based on the CIR baseline model comparison of reflectometric and P2P measurements simultaneously, using two PLC devices at the line ends \cite{Milioudis2012a} and tested on a real network. Before installation, this method requires some pre-deployment experiments to choose the optimal frequency ranges and the number of points for the measurements to be accurate. \textcolor{black}{These setups present a trade-off between setup
complexity and network coverage. As you move from Single-Point to Multi-
Point measurements, setup complexity increases, while network coverage expands.
Specifically, Single-Point measurements have the lowest complexity and the short-
est coverage, while Multi-Point measurements provide broader coverage at the cost
of increased complexity.} Finally, to cover the configuration variations between single and multi-conductor transmission lines, the previous works were extended to simulate and test HIF detection in \acrfull{mtl} configurations. Results showed that a single PLC device installed on one phase could detect faults on the other conductors \cite{Milioudis2015}. 
De Oliveira \emph{et al.} presented an analytical study of different PLC-generated time domain pulses (i.e., \acrfull{hs-ofdm}, impulsive \acrfull{uwb}, and \acrfull{css} for HIF detection \cite{DEOLIVEIRA2022107168}. Results showed that UWB pulses yield a larger number of reflectograms, while HS-OFDM and CSS pulses provide higher reflectogram quality and better resolution in the same time intervals. 
Input impedance measurements were analyzed differently in \cite{Pinomaa2015}. Using a similar fault model as \cite{Milioudis2010}, impedance spectroscopy in the BB-PLC band was employed to detect and locate line faults based on a parameter-adjustable model.

Forstel \emph{et al.} explore the potential of using BB-PLC to detect cable aging in \cite{Forstel2017}. Their proposal uses a physical model that simulates the degradation of service-aged cables due to water treeing to detect changes through CFR observed in PLC signals. The CFR variations are then analyzed and classified using a \acrfull{ml} Support Vector Machine (SVM) algorithm.

The work was extended in \cite{Huo2018a} with the same approach to expand the classification problem to identify the fault condition, whether localized or homogeneous and predict the severity and length of the affected cable part.
Furthermore, since thermal degradation in \acrfull{pilc} cables causes similar effects on the CFR \cite{Huo2018}, the SVM framework of \cite{Forstel2017} was adapted to automatically detect and estimate the severity of thermal degradation faults.
Finally, building on previous findings, a comprehensive ML-based framework using PLC-CFR and \acrfull{jtfdr} was proposed in \cite{Huo2019}. This framework aims to assist in classifying cable aging profiles, assessing the severity of cable degradation, and precisely locating degradation in cases of localized faults.
Additionally, Yang \emph{et al.} proposed a scheme to detect power line faults using CFR measured between two PLMs \cite{Yang2013}. Following the baseline comparison mode, variations of CFR due to changes in power line parameters such as conductivity constant, relative permeability constant, and relative magnetic permeability can lead to the identification of different line faults. 
Liang \emph{et al.} considered detecting ground fault through real-time impedance estimation via CFR \cite{10.3389/fenrg.2023.1103298}. By analyzing the CFR and applying Variational Modal Decomposition (VMD), the proposed method estimates high-frequency input impedance to detect and locate ground faults. ML algorithms are then used to track and identify abnormal states with high sensitivity. Simulation results show effective identification of cable faults, achieving a fault location error of less than 2.13\%.

Passerini \emph{et al.} provided a theoretical study of the high-frequency signal propagation phenomena through power lines, considering the presence of faults \cite{Passerini2019a}. 
The study proposed two models to detect power line anomalies: the chain representation and superposition of effects. These models utilize three key measurements: input admittance, reflection coefficient, and \acrfull{ctf}, based on both reflectometric and end-to-end measurements.
Consequently, these findings were used to design two algorithms for fault detection, classification, and localization \cite{Passerini2019}. OFDM pilots estimate the line input admittance; based on the superposition model, faults are classified as distributed, localized, or load variations. Admittance variations are then used to locate the topology branch at which the fault occurred. 
A similar measurement approach and channel chain representation model was used in \cite{Lehmann2016}. PLC channel was estimated using OFDM pilots, and the CTF $H(f)$ was modeled as a product of the healthy and damaged parts of the line, i.e., $H(f)_{measured}=H(f)_{healthy} \cdot H(f)_{damaged}$ \cite{Lehmann2016}. Therefore, a pre-recorded CFR of the healthy line state is used as a baseline model to extract the CTF damaged component  $H(f)_{damaged}$, which represents an impulse in the CTF time domain representation. 
Hong-shan \emph{et al.} proposed a detection method of aged cable segments based on advanced ML models (i.e., sparse autoencoders and convolutional NNs). The ML models can detect and locate the aging segment based on CTF measurements \cite{https://doi.org/10.1049/smt2.12125}. 
Soft faults were proposed to be detected based on the online transmission coefficient estimation in PLC systems \cite{ABDELKARIM2023}. An algorithm was further developed to deal with complex topology networks by segmenting them into multiple Y-shaped sub-networks. The proposed algorithm was validated on data collected from a real test bench.  

\begin{table*}[!]
\centering
{\color{black} \caption{\textcolor{black}{Summary of Grid Cybersecurity Techniques}}
\label{security_sum}
\setlength\extrarowheight{4pt}
\begin{tabular}{|>\centering m{0.07\linewidth}|>\centering m{0.1\linewidth}|>\centering m{0.15\linewidth}|>\centering m{0.15\linewidth}|>\centering m{0.15\linewidth}|m{0.2\linewidth}|}
\hline
\textbf{References} & \textbf{Application} & \textbf{Technique} & \textbf{Evaluation Metric} & \textbf{Advantages} & \textbf{\* \* \* Disadvantages/Limitations}\\ \hline
 \cite{Lee2019}     & Pairing and
Authentication   &   Generating a bit sequence by indexing noise measurements
based on the maximum absolute voltage value per segment. &    Bit Agreement Rate, and Bit Mismatch Rate &
+ Efficient-cost IoT
device authentication. &
- Limited to local
power line only.\\ 
 \hline
\cite{Yang2020}   & PLS Key Generation  & Extraction of common
bit sequences from PLC signal, by detecting and identifying common spike, udsing K-means clustering & 
 Bit Agreement Rate, Bit Mismatch Rate, Bit Generation Rate, Key Generation Rate, and Key Disagreement Rate & + A powerful security technique. & 
- Increasing the
transmission latency. \newline
- Sensitive to the CSI
estimation error. \\ 
\hline
 \cite{Passerini2020} & PLS Key Generation & Generating a common PLS key, using the CIR, CFR, and input impedance of a given PL link & Spatial Correlation & + A powerful security technique.& 

- Increasing the
transmission latency. \newline
- Sensitive to the PLC channel
estimation error.\\ 
 \hline
 \cite{HernandezFernandez2023}  & Preventing Intrusions in PLC Networks & CIR-based topology change detection and identification technique & Connection Detection Rate & + A powerful security technique.& 
- Increasing the
transmission latency. \newline
- Sensitive to the CSI
estimation error.\\ 
 \hline
  \cite{HernandezFernandez2002PLID}  & PL Authentication & PLC Physical Layer Link Identification with Imperfect CSI & Successful Path Detection Probability & + Accurate identification.
& 
- Multiple PLC nodes
are needed.\\ 
 \hline
\cite{Irfan2023JammingPLC}  & Jamming Detection & Detecting jamming signals using Deep
Learning techniques  & Jamming detection accuracy & + An accurate jamming detection technique.& 
- A high-complexity technique\\ 
 \hline
\end{tabular}}
\end{table*}

\textcolor{black}{In \cite{Rao2011}. Rao \emph{et al.} presented a technique that used smart meter signal strength measurements to discover connection problems, among a set of meters, such as wrong node assignments, rapid signal level drop, and frequent signal level fluctuations. This work  identified 4 metrics that aid smart meter outlier detection: jumpiness score, flatness score, signal drop, and Z-score.} 
\textcolor{black}{Huo \emph{et al.} proposed an anomaly detection and identification method for cable faults based on time series forecasting, without requiring prior domain knowledge. This approach utilizes ML-based time-series forecasting to predict the state of the PLC channel at any given moment, using historical data analysis.}


\subsection{Grid Cybersecurity}

\textcolor{black}{We present in Table \ref{security_sum} a summary of the presented grid cybersecurity works presented in the literature.}\\
\textcolor{black}{In particular,} VoltKey \cite{Lee2019} and PowerKey \cite{Yang2020} are physical layer key generation schemes that use channel parameters in power lines to derive common keys. In a trusted electrical domain, where adversaries lack physical access, pairs or multiple devices can employ these schemes to generate a shared secret key. VolKey uses a protocol to match the time and sampling rate between the pair of devices trying to authenticate. It then generates a bit sequence by indexing noise measurements based on the maximum absolute voltage value per segment.
On the other hand, PowerKey employs \acrfull{emi} spike detection and utilizes K-means clustering to identify common EMI spikes. Unlike VoltKey, this method requires offline processing to identify the frequency windows of the common least correlated EMI spikes for all devices. Key reconciliation is then applied to eliminate any bit errors in the key sequences at the two devices with minimum information exchange.
The extraction of common bit sequences from PLC signals is not limited to EMI, as other sources of common randomness exist in the channel. Passerini \emph{et al.} conducted a study to leverage the reciprocity of PLC channels to generate common information \cite{Passerini2020}. As a result, two key generation schemes were proposed: one that does not require any information exchange and is based on the CIR, and another that uses the transmission matrix of the link, requiring the exchange of the CFR estimates ($H$) and input impedance estimates ($Z_{in}$).

One method for preventing intrusions in PLC networks is to identify unauthorized physical access to the grid by monitoring for changes in the network topology. The authors of \cite{HernandezFernandez2023} propose a CIR-based topology change detection and identification technique. The process begins with channel probing to estimate the different CIRs from the received signals. The second step involves minimizing the channel estimation error by averaging the estimated CIRs \cite{HernandezFernandez2002PLID}. Finally, CIR quantization is performed to generate a unique \acrfull{plid} for each link. This last step is continuously repeated, and the bit Mismatch Rate (BMR) between the previous and new \acrshort{plid} is compared. If the value is above a given threshold it will signal a topology change in that specific link.
Despite being introduced as a topology inference algorithm for indoor PLC, \cite{Zhang2016} presents a similar concept that could also support detection systems.

\cite{Irfan2023JammingPLC} focuses on detecting jamming signals using Deep Learning (DL) techniques and addresses the gap in research on Denial of Service (DoS) attacks in PLC systems. By converting physical layer information (I-Q samples) into images and using these images as input for a Convolutional Neural Networks (CNN), the proposed method achieves jamming detection accuracy higher than 0.99. Unlike traditional solutions that fail when the Bit Error Rate (BER) is zero, the proposed technique remains effective even at distances up to 75 meters from the jammer. 

\color{black}

\section{Discussion}\label{discussion}
 In this section, we provide the insights derived from the study of prior methods and applications in PLC signal analysis, emphasizing the inherent limitations and proposing potential future research directions.

\begin{itemize}
\item \noindent 
The frequency at which measurements are taken and topology estimation algorithms are executed is a key parameter, often influencing the algorithm’s efficiency, complexity, and maintenance of current records. Topology inference frequency can be categorized as either fixed or dynamic. Most reviewed proposals employ a fixed method, although the specific intervals are not detailed. Conversely, \cite{Zhang2016} introduced a technique for dynamic detection and topology inference based on impulsive noise. In order to apply these algorithms in real-world scenarios, dynamic methods are crucial as they enhance efficiency and reduce the complexity of having to re-run topology parameters and graphs unnecessarily.

\item \noindent 
Coverage should be a variable to consider in topology inference models. It can reveal potential blind spots in the network where nodes may not be detected or where faults hinder signal propagation. Combining previous node measurements might provide insights into the unreachable areas or blind spots. However, this approach has not been explored, and such a validation metric has not been addressed in previous contributions.

\item \noindent 
Combining information regarding the type, status, or health of the nodes, along with the electric grid values such as voltage, frequency, and impedance at those points, could further identify blind spots undetected by the topology algorithms. This integrated approach could enhance the accuracy and reliability of the network’s topology inference by highlighting areas that require attention due to potential hidden issues

\item \noindent 
Identifying and locating anomalies is more dependent on the network topology than on the network size or noise levels \cite{Passerini2019}. Most of the reviewed studies focused on single-line or wiretap scenarios, paving the way for further investigation into more characteristic topologies and their impact on the algorithms.

\item \noindent Key generation approaches based on proximity including VoltKey \cite{Lee2019} and PowerKey \cite{Yang2020} offer promising results but assume that PLMs are within a trusted domain isolated from attackers, which might not be realistic. Additionally, these approaches often lack further security considerations, such as insider threats and TEMPEST attacks.

\item \noindent Key distribution processes have not been addressed in PLC systems. The particular channel characteristics of the medium, latency and bandwidth limitations, and topology changes of PLC have not been taken into consideration, and traditional methods are often assumed. Additionally, no research has investigated the use of group or broadcast encryption schemes.

\item \noindent The detection of cyber attacks is another underexplored topic for research. While some studies have addressed the security vulnerabilities of protocols in PLC environments \cite{6201308, seijo2017cybersecurity, 9628606}, the jamming detection solution \cite{Irfan2023JammingPLC} was the only one found using physical layer information to address these issues.  

\item \noindent Topology inferring algorithms, particularly those focusing on topology changes or providing quick and lightweight re-estimation methods, such as \cite{Zhang2016} and \cite{HernandezFernandez2023}, could easily be integrated into intrusion detection systems. These algorithms can enhance their detection capabilities by identifying unauthorized modifications in the network topology, which could indicate malicious activities.

\end{itemize}

\color{black}
\section{Conclusion}\label{conclusion}

Existing surveys of PLC technology have predominantly concentrated on the communications aspect or specific applications, with none focusing on the grid information inference component. This unique capability of PLC systems to infer information about the medium while transmitting data makes this technology particularly interesting and efficient, as it allows for stacking different functionalities within the same device. Despite this advantage, there are no comprehensive reviews that focus on the grid information inference component, studying its limitations, pinpointing research gaps, and providing ideas for future research directions.
In this paper, we classified the research contributions related to PLC grid information inference, studied their techniques, and evaluated their merits and limitations. We drew conclusions about the current status of the field, suggested improvement vectors, and highlighted existing research gaps in this still uncharted domain.




\section*{Acknowledgement}
This publication is supported by Iberdrola S.A. as part of its innovation department research studies. Its contents are solely the responsibility of the authors and do not necessarily represent the official views of Iberdrola Group.

\balance
\bibliographystyle{main}
\bibliography{main}

\begin{thebibliography}{10}

\bibitem{7467440}
Cristina Cano, Alberto Pittolo, David Malone, Lutz Lampe, Andrea~M. Tonello, and Anand~G. Dabak,
\newblock ``{State of the Art in Power Line Communications: From the Applications to the Medium},''
\newblock {\em IEEE Journal on Selected Areas in Communications}, vol. 34, no. 7, pp. 1935--1952, 2016.

\bibitem{YIGIT2014}
Melike Yigit, V.~Cagri Gungor, Gurkan Tuna, Maria Rangoussi, and Etimad Fadel,
\newblock ``{Power Line Communication Technologies for Smart Grid Applications: A Review of Advances and Challenges},''
\newblock {\em Computer Networks}, vol. 70, pp. 366--383, 2014.

\bibitem{Yoldas2017EnhancingSGwithMIC}
Yeliz Yoldaş, Ahmet Önen, S.M. Muyeen, Athanasios~V. Vasilakos, and İrfan Alan,
\newblock ``Enhancing smart grid with microgrids: Challenges and opportunities,''
\newblock {\em Renewable and Sustainable Energy Reviews}, vol. 72, pp. 205--214, 2017.

\bibitem{AGUERO2018}
Julio Romero~Aguero and Amin Khodaei,
\newblock ``{Grid Modernization, DER Integration \& Utility Business Models - Trends \& Challenges},''
\newblock {\em IEEE Power and Energy Magazine}, vol. 16, no. 2, pp. 112--121, 2018.

\bibitem{machowski2020power}
Jan Machowski, Zbigniew Lubosny, Janusz~W Bialek, and James~R Bumby,
\newblock {\em {Power System Dynamics: Stability and Control}},
\newblock John Wiley \& Sons, 2020.

\bibitem{1519722}
F.F. Wu, K.~Moslehi, and A.~Bose,
\newblock ``{Power System Control Centers: Past, Present, and Future},''
\newblock {\em Proceedings of the IEEE}, vol. 93, no. 11, pp. 1890--1908, 2005.

\bibitem{5503917}
Victor O.~K. Li, Felix~F. Wu, and Jin Zhong,
\newblock ``{Communication Requirements for Risk-Limiting Dispatch in Smart Grid},''
\newblock in {\em 2010 IEEE International Conference on Communications Workshops}, 2010, pp. 1--5.

\bibitem{SAIDU20223192}
Mustapha~Muhammad Saidu, Shiva~Pujan Jaiswal, Kuldeep Jayaswal, Shibamay Mitra, and Vikas~Singh Bhadoria,
\newblock ``{A Survey on: Automation of Micro Grid and Micro Distributed Generation},''
\newblock {\em Materials Today: Proceedings}, vol. 49, pp. 3192--3196, 2022,
\newblock National Conference on Functional Materials: Emerging Technologies and Applications in Materials Science.

\bibitem{Guelpa2021DRSurvey}
Elisa Guelpa and Vittorio Verda,
\newblock ``Demand response and other demand side management techniques for district heating: A review,''
\newblock {\em Energy}, vol. 219, pp. 119440, 2021.

\bibitem{RIVAS2020}
Angel~Esteban {Labrador Rivas} and Taufik Abrão,
\newblock ``{Faults in Smart Grid Systems: Monitoring, Detection and Classification},''
\newblock {\em Electric Power Systems Research}, vol. 189, pp. 106602, 2020.

\bibitem{Kababji2024DesignPLC}
Ayman Al-Kababji, José~Miguel Sanz-Alcaine, Alfredo Sanz, and Javier~Hernandez Fernandez,
\newblock ``Advancing plc network analysis: The design of a comprehensive plc media characterization system,''
\newblock in {\em 2024 IEEE 8th Energy Conference (ENERGYCON)}, 2024, pp. 1--6.

\bibitem{passerini2017NetworkSensing}
Federico Passerini and Andrea~M. Tonello,
\newblock ``{Full Duplex Power Line Communication Modems for Network Sensing},''
\newblock in {\em 2017 IEEE International Conference on Smart Grid Communications (SmartGridComm)}, 2017, pp. 213--217.

\bibitem{HernandezFernandez2023performancePLS}
Javier~Hernandez Fernandez, Aymen Omri, and Roberto Di~Pietro,
\newblock ``Performance analysis of physical layer security in power line communication networks,''
\newblock in {\em 2023 IEEE Symposium on Computers and Communications (ISCC)}, 2023, pp. 777--782.

\bibitem{Henkel2020}
Werner Henkel, Abderraheem~M. Turjman, Hayoung Kim, and Hisham~K.H. Qanadilo,
\newblock ``{Common Randomness for Physical-Layer Key Generation in Power-Line Transmission},''
\newblock {\em IEEE International Conference on Communications}, vol. 2020-June, pp. 1--6, 2020.

\bibitem{HernandezFernandez2023}
Javier {Hernandez Fernandez}, Aymen Omri, and Roberto {Di Pietro},
\newblock ``{Power Grid Surveillance: Topology Change Detection System Using Power Line Communications},''
\newblock {\em International Journal of Electrical Power \& Energy Systems}, vol. 145, no. September 2022, pp. 108634, 2023.

\bibitem{Furse2006}
Cynthia Furse, You Chung, Chet Lo, and Praveen Pendayala,
\newblock ``{A Critical Comparison Of Reflectometry Methods For Location Of Wiring Faults},''
\newblock {\em Smart Structures and Systems}, vol. 2, no. 1, pp. 25--46, 2006.

\bibitem{Shi2010}
Qinghai Shi, Uwe Troeltzsch, and Olfa Kanoun,
\newblock ``{Detection and Localization of Cable Faults by Time and Frequency Domain Measurements},''
\newblock {\em 2010 7th International Multi-Conference on Systems, Signals and Devices, SSD-10}, , no. 2, 2010.

\bibitem{Passerini2018}
Federico Passerini and Andrea~M. Tonello,
\newblock ``{Power Line Communications for Grid Discovery and Diagnostics},''
\newblock {\em Encyclopedia of Wireless Networks}, pp. 1--6, 2018.

\bibitem{PLCSurvey}
Jean Paul~A. Yaacoub, Javier {Hernandez Fernandez}, Hassan~N. Noura, and Ali Chehab,
\newblock ``Security of power line communication systems: Issues, limitations and existing solutions,''
\newblock {\em Computer Science Review}, vol. 39, pp. 100331, 2021.

\bibitem{Galli2011}
Stefano Galli, Anna Scaglione, and Zhifang Wang,
\newblock ``{For The Grid and Through The Grid: The Role of Power Line Communications in The Smart Grid},''
\newblock {\em Proceedings of the IEEE}, vol. 99, no. 6, pp. 998--1027, 2011.

\bibitem{PLSinPLC_Book_2024}
Javier {Hernandez Fernandez}, Aymen Omri, and Roberto {Di Pietro},
\newblock {\em Physical Layer Security in Power Line Communications: Fundamentals, Models and Applications},
\newblock Springer, 2024.

\bibitem{hallack2016PLChomeandIndch7}
G.~Hallak and G.~Bumiller,
\newblock {\em PLC for Home and Industry Automation}, chapter~7, pp. 449--472,
\newblock John Wiley \& Sons, Ltd, 2016.

\bibitem{pinto2015}
Freddy~A. Pinto-Benel and Fernando Cruz-Roldán,
\newblock ``{2-ASCET for Broadband Multicarrier Transmission over In-home and In-vehicle Power Line Networks},''
\newblock in {\em 2015 IEEE 18th International Conference on Intelligent Transportation Systems}, 2015, pp. 1351--1356.

\bibitem{antoniali2011}
Massimo Antoniali, Andrea~M. Tonello, Matteo Lenardon, and Andrea Qualizza,
\newblock ``{Measurements and Analysis of PLC Channels in A Cruise Ship},''
\newblock in {\em 2011 IEEE International Symposium on Power Line Communications and Its Applications}, 2011, pp. 102--107.

\bibitem{degardin2013}
V.~Degardin, I.~Junqua, M.~Lienard, P.~Degauque, and S.~Bertuol,
\newblock ``{Theoretical Approach to the Feasibility of Power-Line Communication in Aircrafts},''
\newblock {\em IEEE Transactions on Vehicular Technology}, vol. 62, no. 3, pp. 1362--1366, 2013.

\bibitem{SHARMA2017}
Konark Sharma and Lalit~Mohan Saini,
\newblock ``{Power-Line Communications for Smart Grid: Progress, Challenges, Opportunities and Status},''
\newblock {\em Renewable and Sustainable Energy Reviews}, vol. 67, pp. 704--751, 2017.

\bibitem{ergodic2022}
Javier~Hernandez Fernandez, Luis Lacasa, Aymen Omri, Alfredo Sanz, and Munther~E. Koborsi,
\newblock ``Ergodic capacity analysis of ofdm-based nb-plc systems,''
\newblock in {\em 2022 24th International Conference on Advanced Communication Technology (ICACT)}, 2022, pp. 399--405.

\bibitem{en13123098}
Bilal Masood, M.~Arif Khan, Sobia Baig, Guobing Song, Ateeq~Ur Rehman, Saif~Ur Rehman, Rao~M. Asif, and Muhammad~Babar Rasheed,
\newblock ``{Investigation of Deterministic, Statistical and Parametric NB-PLC Channel Modeling Techniques for Advanced Metering Infrastructure},''
\newblock {\em Energies}, vol. 13, no. 12, 2020.

\bibitem{Liang2019}
Dong Liang, Huashan Guo, and Tao Zheng,
\newblock ``{Real-Time Impedance Estimation for Power Line Communication},''
\newblock {\em IEEE Access}, vol. 7, pp. 88107--88115, 2019.

\bibitem{Lehmann2016}
Andreas~M. Lehmann, Katrin Raab, Florian Gruber, Erik Fischer, Ralf Muller, and Johannes~B. Huber,
\newblock ``{A Diagnostic Method for Power Line Networks by Channel Estimation of PLC Devices},''
\newblock {\em 2016 IEEE International Conference on Smart Grid Communications, SmartGridComm 2016}, pp. 320--325, 2016.

\bibitem{Pappu2018}
Satya~Jayadev Pappu, Nirav Bhatt, Ramkrishna Pasumarthy, and Aravind Rajeswaran,
\newblock ``{Identifying Topology of Low Voltage Distribution Networks Based on Smart Meter Data},''
\newblock {\em IEEE Transactions on Smart Grid}, vol. 9, no. 5, pp. 5113--5122, 2018.

\bibitem{Diwold2015}
Konrad Diwold, Matthias Stifter, and Paul Zehetbauer,
\newblock ``{Network And Feeder Assignment Of Smart Meters Based On Communication And Measurement Data},''
\newblock {\em Proceedings - 2015 International Symposium on Smart Electric Distribution Systems and Technologies, EDST 2015}, , no. September, pp. 541--546, 2015.

\bibitem{Huo2019}
Yinjia Huo, Gautham Prasad, Lazar Atanackovic, Lutz Lampe, and Victor~C.M. Leung,
\newblock ``{Cable Diagnostics with Power Line Modems for Smart Grid Monitoring},''
\newblock {\em IEEE Access}, vol. 7, pp. 60206--60220, 2019.

\bibitem{Milioudis2012a}
Apostolos~N. Milioudis, Georgios~T. Andreou, and Dimitrios~P. Labridis,
\newblock ``{Enhanced Protection Scheme for Smart Grids Using Power Line Communications Techniques-Part II: Location of High Impedance Fault Position},''
\newblock {\em IEEE Transactions on Smart Grid}, vol. 3, no. 4, pp. 1631--1640, 2012.

\bibitem{Milioudis2015}
Apostolos~N. Milioudis, Georgios~T. Andreou, and Dimitris~P. Labridis,
\newblock ``{Detection and Location of High Impedance Faults in Multiconductor Overhead Distribution Lines Using Power Line Communication Devices},''
\newblock {\em IEEE Transactions on Smart Grid}, vol. 6, no. 2, pp. 894--902, 2015.

\bibitem{Huo2018a}
Yinjia Huo, Gautham Prasad, Lazar Atanackovic, Lutz Lampe, and Victor~C.M. Leung,
\newblock ``{Grid Surveillance and Diagnostics Using Power Line Communications},''
\newblock {\em 2018 IEEE International Symposium on Power Line Communications and its Applications, ISPLC 2018}, pp. 1--6, 2018.

\bibitem{Huo2018}
Yinjia Huo, Gautham Prasad, Lutz Lampe, and Victor~C.M. Leung,
\newblock ``{Cable Health Monitoring in Distribution Networks using Power Line Communications},''
\newblock {\em 2018 IEEE International Conference on Communications, Control, and Computing Technologies for Smart Grids, SmartGridComm 2018}, pp. 1--6, 2018.

\bibitem{Passerini2019a}
Federico Passerini and Andrea~M. Tonello,
\newblock ``{Smart Grid Monitoring Using Power Line Modems: Effect of Anomalies on Signal Propagation},''
\newblock {\em IEEE Access}, vol. 7, pp. 27302--27312, 2019.

\bibitem{Passerini2019}
Federico Passerini and Andrea~M. Tonello,
\newblock ``{Smart Grid Monitoring Using Power Line Modems: Anomaly Detection and Localization},''
\newblock {\em IEEE Transactions on Smart Grid}, vol. 10, no. 6, pp. 6178--6186, 2019.

\bibitem{LABRADORRIVAS2020106602}
Angel~Esteban {Labrador Rivas} and Taufik Abrão,
\newblock ``Faults in smart grid systems: Monitoring, detection and classification,''
\newblock {\em Electric Power Systems Research}, vol. 189, pp. 106602, 2020.

\bibitem{Rao2011}
Rakesh Rao, Srinivas Akella, and Gokhan Guley,
\newblock ``{Power Line Carrier (PLC) Signal Analysis of Smart Meters for Outlier Detection},''
\newblock {\em 2011 IEEE International Conference on Smart Grid Communications, SmartGridComm 2011}, pp. 291--296, 2011.

\bibitem{10.1007/978-3-319-23609-4_8}
Alberto Pittolo and Andrea~M. Tonello,
\newblock ``Physical layer security in power line communication networks,''
\newblock in {\em Physical and Data-Link Security Techniques for Future Communication Systems}, Marco Baldi and Stefano Tomasin, Eds., Cham, 2016, pp. 125--144, Springer International Publishing.

\bibitem{Ahmed2012}
Mohamed~Osama Ahmed and Lutz Lampe,
\newblock ``{Power Line Network Topology Inference Using Frequency Domain Reflectometry},''
\newblock {\em IEEE International Conference on Communications}, , no. 3, pp. 3419--3423, 2012.

\bibitem{Zhang2016}
Chao Zhang, Xu~Zhu, Yi~Huang, and Gan Liu,
\newblock ``{High-Resolution and Low-Complexity Dynamic Topology Estimation for Plc Networks Assisted by Impulsive Noise Source Detection},''
\newblock {\em IET Communications}, vol. 10, no. 4, pp. 443--451, 2016.

\bibitem{Ulrich2015}
Michael Ulrich and Bin Yang,
\newblock ``{Inference of Wired Network Topology Using Multipoint Reflectometry},''
\newblock pp. 1965--1969, 2015.

\bibitem{Ahmed2013}
Mohamed~O. Ahmed and Lutz Lampe,
\newblock ``{Power Line Communications for Low-Voltage Power Grid Tomography},''
\newblock {\em IEEE Transactions on Communications}, vol. 61, no. 12, pp. 5163--5175, 2013.

\bibitem{Ma2015}
Xu~Ma, Fang Yang, Wenbo Ding, and Jian Song,
\newblock ``{Topology Reconstruction for Power Line Network Based on Bayesian Compressed Sensing},''
\newblock {\em 2015 IEEE International Symposium on Power Line Communications and Its Applications, ISPLC 2015}, pp. 119--124, 2015.

\bibitem{Lampe2013}
Lutz Lampe and Mohamed~O. Ahmed,
\newblock ``{Power Grid Topology Inference Using Power Line Communications},''
\newblock {\em 2013 IEEE International Conference on Smart Grid Communications, SmartGridComm 2013}, pp. 336--341, 2013.

\bibitem{Erseghe2013}
Tomaso Erseghe, Stefano Tomasin, and Alberto Vigato,
\newblock ``{Topology Estimation for Smart Micro Grids via Powerline Communications},''
\newblock {\em IEEE Transactions on Signal Processing}, vol. 61, no. 13, pp. 3368--3377, 2013.

\bibitem{Passerini2017}
Federico Passerini and Andrea~M. Tonello,
\newblock ``{On the Exploitation of Admittance Measurements for Wired Network Topology Derivation},''
\newblock {\em IEEE Transactions on Instrumentation and Measurement}, vol. 66, no. 3, pp. 374--382, 2017.

\bibitem{Aouichak2017}
Ismail Aouichak, Kassim Khalil, Imene Elfeki, Jean~Charles {Le Bunetel}, and Yves Raingeaud,
\newblock ``{Topology Identification Method for Unknown Indoor PLC Home Networks},''
\newblock {\em 2017 International Symposium on Electromagnetic Compatibility - EMC EUROPE 2017, EMC Europe 2017}, pp. 6--9, 2017.

\bibitem{Costabeber2011}
Alessandro Costabeber, Paolo Tenti, Tomaso Erseghe, Stefano Tomasin, and Paolo Mattavelli,
\newblock ``{Distributed Control of Smart Microgrids by Dynamic Grid Mapping},''
\newblock {\em IECON Proceedings (Industrial Electronics Conference)}, pp. 1323--1328, 2011.

\bibitem{Wang2019}
Xueliang Wang, Xun Dai, Huizhu Xian, Weiqiong Song, Baiyu Gao, Shuming Xu, Lin Pang, and Shuai Guo,
\newblock ``{Fault Localization for Power Line Communications with Topology Inference},''
\newblock {\em Proceedings - 2019 International Conference on Intelligent Computing, Automation and Systems, ICICAS 2019}, pp. 85--90, 2019.

\bibitem{Marron2013}
Laura Marron, Xabier Osorio, Asier Llano, Aitor Arzuaga, and Alberto Sendin,
\newblock ``{Low Voltage Feeder Identification for Smart Grids with Standard Narrowband PLC Smart Meters},''
\newblock {\em ISPLC 2013 - 2013 IEEE 17th International Symposium on Power Line Communications and Its Applications, Proceedings}, pp. 120--125, 2013.

\bibitem{Mlynek2016}
Petr Mlynek, Radek Fujdiak, and Jiri Misurec,
\newblock ``{Power Line Topology Prediction Using Time Domain Reflectometry},''
\newblock {\em 2016 39th International Conference on Telecommunications and Signal Processing, TSP 2016}, pp. 199--202, 2016.

\bibitem{Ravishankar2016}
S.~Ravishankar and R.~Arjun,
\newblock ``{A Hybrid Method for Physical and Power Line Loop Topology Estimation Using A Broadband Modem},''
\newblock {\em IEEE Radio and Wireless Symposium, RWS}, vol. 2016-March, pp. 99--102, 2016.

\bibitem{Ravishankar2012LineTE}
S.~Ravishankar and M.~Bharathi,
\newblock ``{Line Topology Estimation of Indoor Power Lines Using Multipoint Single Ended Loop Testing},''
\newblock 2012.

\bibitem{Ahmed2012a}
Mohamed~Osama Ahmed and Lutz Lampe,
\newblock ``{Parametric and Nonparametric Methods for Power Line Network Topology Inference},''
\newblock {\em 2012 IEEE International Symposium on Power Line Communications and Its Applications, ISPLC 2012}, pp. 274--279, 2012.

\bibitem{Milioudis2010}
A.~N. Milioudis, G.~T. Andreou, and D.~P. Labridis,
\newblock ``{High Impedance Fault Detection Using Power Line Communication Techniques},''
\newblock {\em Proceedings of the Universities Power Engineering Conference}, pp. 1--6, 2010.

\bibitem{Milioudis2011}
A.~N. Milioudis, G.~T. Andreou, and D.~P. Labridis,
\newblock ``{High Impedance Fault Evaluation Using Narrowband Power Line Communication Techniques},''
\newblock {\em 2011 IEEE PES Trondheim PowerTech: The Power of Technology for a Sustainable Society, POWERTECH 2011}, pp. 1--6, 2011.

\bibitem{Milioudis2012}
Apostolos~N. Milioudis, Georgios~T. Andreou, and Dimitrios~P. Labridis,
\newblock ``{Enhanced Protection Scheme for Smart Grids Using Power Line Communications Techniques - Part I: Detection of High Impedance Fault Occurrence},''
\newblock {\em IEEE Transactions on Smart Grid}, vol. 3, no. 4, pp. 1621--1630, 2012.

\bibitem{DEOLIVEIRA2022107168}
Lucas~Giroto {de Oliveira}, Mateus de~L.~Filomeno, Luiz~Fernando Colla, H.~{Vincent Poor}, and Moisés~V. Ribeiro,
\newblock ``{Analysis of Typical PLC Pulses for Sensing High-Impedance Faults Based on Time-Domain Reflectometry},''
\newblock {\em International Journal of Electrical Power \& Energy Systems}, vol. 135, pp. 107168, 2022.

\bibitem{Forstel2017}
Lena Forstel and Lutz Lampe,
\newblock ``{Grid Diagnostics: Monitoring Cable Aging Using Power Line Transmission},''
\newblock {\em 2017 IEEE International Symposium on Power Line Communications and its Applications, ISPLC 2017}, pp. 1--6, 2017.

\bibitem{Yang2013}
Fang Yang, Wenbo Ding, and Jian Song,
\newblock ``{Non-Intrusive Power Line Quality Monitoring Based On Power Line Communications},''
\newblock {\em ISPLC 2013 - 2013 IEEE 17th International Symposium on Power Line Communications and Its Applications, Proceedings}, pp. 191--196, 2013.

\bibitem{Huo2022}
Yinjia Huo, Gautham Prasad, Lutz Lampe, and Victor Leung,
\newblock ``{Power Line Communication and Sensing Using Time Series Forecasting},''
\newblock {\em Sensors}, vol. 22, no. 14, pp. 1--19, 2022.

\bibitem{Pinomaa2015}
Antti Pinomaa, Jero Ahola, Antti Kosonen, and Tero Ahonen,
\newblock ``{Diagnostics Of Low-Voltage Power Cables By Using Broadband Impedance Spectroscopy},''
\newblock {\em 2015 17th European Conference on Power Electronics and Applications, EPE-ECCE Europe 2015}, 2015.

\bibitem{10.3389/fenrg.2023.1103298}
Dong Liang, Kaiwen Zhang, Song Ge, Yang Wang, and Deyi Wang,
\newblock ``{A Novel Fault Monitoring Method Based on Impedance Estimation of Power Line Communication Equipment},''
\newblock {\em Frontiers in Energy Research}, vol. 11, 2023.

\bibitem{https://doi.org/10.1049/smt2.12125}
Zhao Hong-shan, Guo Xiao-mei, Ma~Li-bo, Wang Yan, Sun Cheng-yan, and Chang Jie-ying,
\newblock ``{A Hierarchical Diagnosis Method of Cable Aged Segment Based on Transfer Function},''
\newblock {\em IET Science, Measurement \& Technology}, vol. 16, no. 9, pp. 512--522, 2022.

\bibitem{ABDELKARIM2023}
Abdel~Karim {Abdel Karim}, M.~Amine Atoui, Virginie Degardin, Pierre Laly, and Vincent Cocquempot,
\newblock ``Bus network decomposition for fault detection and isolation through power line communication,''
\newblock {\em ISA Transactions}, 2023.

\bibitem{Lee2019}
Kyuin Lee, Neil Klingensmith, Suman Banerjee, and Younghyun Kim,
\newblock ``{VoltKey: Continuous Secret Key Generation Based on Power Line Noise for Zero-Involvement Pairing and Authentication},''
\newblock {\em Proceedings of the ACM on Interactive, Mobile, Wearable and Ubiquitous Technologies}, vol. 3, no. 3, pp. 1--26, 2019.

\bibitem{Yang2020}
Fangfang Yang, Mohammad~A. Islam, and Shaolei Ren,
\newblock ``{PowerKey: Generating Secret Keys from Power Line Electromagnetic Interferences},''
\newblock {\em Lecture Notes in Computer Science (including subseries Lecture Notes in Artificial Intelligence and Lecture Notes in Bioinformatics)}, vol. 12570 LNCS, pp. 354--370, 2020.

\bibitem{Passerini2020}
Federico Passerini and Andrea~M. Tonello,
\newblock ``{Secure PHY Layer Key Generation in the Asymmetric Power Line Communication Channel},''
\newblock {\em Electronics (Switzerland)}, vol. 9, no. 4, 2020.

\bibitem{HernandezFernandez2002PLID}
Javier Hernandez~Fernandez, Aymen Omri, and Roberto Di~Pietro,
\newblock ``{PLC Physical Layer Link Identification with Imperfect Channel State Information},''
\newblock {\em Energies}, vol. 15, no. 16, 2022.

\bibitem{Irfan2023JammingPLC}
Muhammad Irfan, Aymen Omri, Javier~Hernandez Fernandez, Savio Sciancalepore, and Gabriele Oligeri,
\newblock ``Jamming detection in power line communications leveraging deep learning techniques,''
\newblock in {\em 2023 International Symposium on Networks, Computers and Communications (ISNCC)}, 2023, pp. 1--6.

\bibitem{6201308}
Bernd Hirschler and Albert Treytl,
\newblock ``{Internet Protocol Security and Power Line Communication},''
\newblock in {\em 2012 IEEE International Symposium on Power Line Communications and Its Applications}, 2012, pp. 102--107.

\bibitem{seijo2017cybersecurity}
Miguel Seijo~Sim{\'o}, Gregorio L{\'o}pez~L{\'o}pez, and Jos{\'e}~Ignacio Moreno~Novella,
\newblock ``{Cybersecurity Vulnerability Analysis of The PLC Prime Standard},''
\newblock {\em Security and Communication Networks}, vol. 2017, 2017.

\bibitem{9628606}
Emmanuel~C. Uwaezuoke and Theo~G. Swart,
\newblock ``{Network Attack Analysis of an Indoor Power Line Communication Network},''
\newblock in {\em 2021 IEEE International Symposium on Power Line Communications and its Applications (ISPLC)}, 2021, pp. 96--101.

\end{thebibliography}

\end{document}